\documentclass[amsfonts,aps,preprint]{revtex4}

\begin{document}

\title{GENERAL UNITARY $SU(1,1)$ TFD FORMULATION}

\author{ M. C. B. Abdalla{\footnote {mabdalla@ift.unesp.br}}
and A. L. Gadelha{\footnote {gadelha@ift.unesp.br}}}

\affiliation{Instituto de F\'{\i}sica Te\'{o}rica, Unesp, Pamplona 145,
S\~{a}o Paulo, SP, 01405-900, Brazil }

\begin{abstract}
In this work we discuss a generalization for the thermal Bogoliubov transformation
in the context of a hermitian general $SU(1,1)$ transformation generator. 
The TFD tilde conjugation rules are redefined using an appropriated Tomita-Takesaki
modular operator.
\end{abstract}

\maketitle

Since the original formulation of Thermo Field Dynamic (TFD) was presented
{\cite{lapa,tu}}, it has been used to deal with a considerable number of
physical systems. From the superconductivity {\cite{boo}} to
string theory {\cite{yl1,yl2,yl3,yl4,yl5,yl6,fn1,fn2,fn3,fn4,fn5,hf,fn6,cei}},
many branches of physics were approached by that formulation and it contributes 
to shed some light over physics {\it fenomena } as the Casimir effect {\cite{johe}},
quantum dissipation {\cite{vitt}, spontaneous symmetry breakdown {\cite{um1}}, black
hole radiation {\cite{isrra,tan}, quantum information theory {\cite{admit}} and
coherent states at finite temperature {\cite{mann}. The latter is closely related
with the systematic construction of thermal boundary states interpreted as
$D$-branes at finite temperature {\cite{ivv,agv1,agv2,agv3,agv4}}.
Par\-al\-lel with this development, the formulation itself has been subject of
intense study as in the seminal work by Ojima {\cite{oji}} that shows the
equivalence between TFD and the Gelfand-Naimark-Segal (GNS) {\cite{emch}}
construction, more precisely with the Haag-Hugenholtz-Winnink (HHW) algebraic
formulation {\cite{lvw}}. The intrinsic structure of the formulation was explored
and a thermal algebra arose {\cite{ad1}}, contributing, for instance, to a
better understanding of the doubling of degrees of freedom as well as a systematic
study of the Galilei and Poincar\'{e} groups {\cite{ad2}}, leading to many results
as the relativistic Wigner density for Klein-Gordon and Dirac fields. 

The algebraic nature of the formalism provides a powerful and suitable approach for
the understanding of microscopic statistical properties, mainly when one makes the
choice to deal with the canonical quantum formulation. In that context, the TFD is
the only possible construction to lead systems at zero temperature to a temperature
dependent one {\cite{chume}}.

In TFD, the thermal effects appear through the mixing between elements of the
original physical space and elements of an auxiliary space, identical to the
original one, in such a way that the total (doubled) space is given by
$\widehat{{\cal H}}={\cal H} \otimes \widetilde{\cal H}$. That mixing is promoted
by a specific Bogoliubov transformation called thermal Bogoliubov transformation
{\cite{ume}}. This transformation must be canonical in the sense that the algebra
of the original system remains the same, keeping its dynamic. The temperature
dependence comes from the transformation's parameters.

Effectively, there are three generators that assure the transformation cited above.
These generators satisfy a $su(1,1)$ algebra, in the case of a bosonic system,
and a $su(2)$ algebra for fermions {\cite{chume}. A general formulation using those
three generators was developed in such a way that, the transformation operator, whose
generator is a combination of the three generators, commutes with the Tomita-Takesaki
modular operator, which in turn defines the map between the physical and auxiliary
systems. The tilde conjugation rules {\cite{ume,lvw}} now remain unchanged. This
development uses a non-unitary transformation, in the finite volume limit, to lead
the system to a finite temperature and therefore, the elements of the transformed
Fock space must be redefined.

The general formulation mentioned above is not the only one possible. Recently, a
general unitary formulation was presented {\cite{agv2}}, following the one proposed
by Elmfors and Umezawa {\cite{elmume}}. In that work the authors introduced the
possibility of constructing a general $SU(1,1)$ unitary transformation, at the finite
volume limit, in a system with an infinite degrees of freedom, namely bosonic string
theory, responsible for leading the system from zero to finite temperature. That
construction presents an apparent inconvenience: the non invariance of the thermal
vacuum under tilde conjugation rules. In Ref. {\cite{agv2}} it was argued that a possible
ambiguity in the choice of thermal states emerges. However, as pointed out in
Ref. {\cite{elmume}, it reflects the fact that under the construction, the
Tomita-Takesaki modular operator does not commute with the transformation operator
and this gives rise to an apparent ambiguity, for example, in the thermal vacuum choice,
as presented in {\cite{agv2}}. In this way, in order to keep the structure of the
TFD formulation and to avoid ambiguities, the tilde conjugation rules must be
redefined in the transformed space {\cite{elmume}}.

In this letter we present the general $SU(1,1)$ unitary transformation as well as the
new (transformed) Tomita-Takesaki modular operator that maintain the TFD structure
and eliminates any possible ambiguities coming from this general formulation. In the
following, we consider just one degree of freedom, but our results can be easily extended
to a field theory, as is the case in Ref. \cite{agv2}.

Consider the bosonic creation and annihilation operators $A^{\dagger}$ and $A$,
respectively, and their associated partners of the auxiliary space, denoted by a tilde.
These two sets of operators satisfy an oscillator-like algebra. The generators for the
thermal transformation are given by
\begin{equation}
g_{1}\left( \theta \right) =\theta _{1}\left( A\widetilde{A}+\widetilde{A}
^{\dagger }A^{\dagger }\right) ,\quad g_{2}\left( \gamma \right) =i\theta
_{2}\left( A\widetilde{A}-\widetilde{A}^{\dagger }A^{\dagger }\right) , 
\quad g_{3}\left( \theta \right) =\theta _{3}\left( A^{\dagger }A+\widetilde{A}
\widetilde{A}^{\dagger }\right) , 
\end{equation}
with $\theta _{1}$, $\theta _{2}$ e $\theta _{3}$ $\in {\mathbb R}$. As it was said
before, elements of the original space are mapped on those of the auxiliary tilde space,
and {\it vice-versa}, using the Tomita-Takesaki modular operator, $J$, as follows
\begin{equation}
\widetilde{A}=JAJ,\qquad J\widetilde{A}J=A,
\end{equation}
where we have used the property $JJ=1$ {\cite{lvw}}. These generators can be combined
in order that they compose the total generator of the general $SU(1,1)$ unitary
transformation as the following 
\begin{equation}
G\left( \theta \right) =\sum_{i} g_{i}\left( \theta \right) . 
\end{equation}
Explicitly we have
\begin{equation}
G\left( \theta \right) =\lambda _{1}\left( \theta \right) \widetilde{A}
^{\dagger }A^{\dagger }-\lambda _{2}\left( \theta \right) A\widetilde{A}
+\lambda _{3}\left( \theta \right) \left( A^{\dagger }A+\widetilde{A}
\widetilde{A}^{\dagger }\right) , 
\label{ge}
\end{equation}
with the $\lambda$'s coefficients given by
\begin{equation}
\lambda _{1}\left( \theta \right) =\theta _{1}-i\theta _{2},\qquad \lambda
_{2}\left( \theta \right) =-\lambda _{1}^{*},\qquad \lambda _{3}\left(
\theta \right) =\theta _{3}. 
\end{equation}
Observing the generator expressed in equation (\ref{ge}), we note that
\begin{equation}
G\left( \theta \right)=G^{\dagger}\left( \theta \right),
\qquad \widetilde{G}\left( \theta \right)=JG\left( \theta \right)J=
G^{*}\left( \theta \right) \neq - G\left( \theta \right).
\label{gprop}
\end{equation}
The first property exprimes that $G(\theta)$ is hermitian and assures that
in the finite volume limit, the transformation is unitary. The second property
indicates that the tilde substitution rules are not preserved and so, must be
redefined. This is more clear when we observe the expressions for the transformed
operators and the thermal state. As it was commented, the thermal operators are
obtained by executing the thermal Bogoliubov transformation which is canonical.
Using the notation $\breve{A}(\theta)$, indicating the new mapping rules in the
transformed space, similar to the tilde rules, the transformation for operators
must take the following form
\begin{eqnarray}
\left( 
\begin{array}{c}
A(\theta) \\ 
\breve{A}^{\dagger}(\theta)
\end{array}
\right) &=&e^{-iG}\left( 
\begin{array}{c}
A \\ 
\widetilde{A}^{\dagger }
\end{array}
\right) e^{iG}={\mathbb B}\left( 
\begin{array}{c}
A \\ 
\widetilde{A}^{\dagger }
\end{array}
\right) , 
\label{tbu}
\\
\left( 
\begin{array}{cc}
A^{\dagger}(\theta) & -\breve{A}(\theta)
\end{array}
\right) &=&\left( 
\begin{array}{cc}
A^{\dagger } & -\widetilde{A}
\end{array}
\right) {\mathbb B}^{-1},
\label{tbui}
\end{eqnarray}
where the $SU(1,1)$ matrix transformation is given by
\begin{eqnarray}
{\mathbb B}=\left( 
\begin{array}{cc}
u & v \\ 
v^{*} & u^{*}
\end{array}
\right) ,\qquad uu^{*}-vv^{*}=1, 
\end{eqnarray}
with elements
\begin{equation}
u=\cosh \left( i\Lambda \right) +\frac{\lambda _{3}}{\Lambda }
\mbox{senh}\left(i\Lambda \right) , \quad
v=\frac{\lambda _{1}}{\Lambda }\mbox{sinh}\left( i\Lambda \right) , 
\end{equation}
where $\Lambda$ is defined by the following relation
\begin{equation}
\Lambda ^{2}=\lambda _{1}\lambda _{2}+\lambda _{3}^{2}. 
\end{equation}
The $SU(1,1)$ matrix transformation can be written in terms of
the $\lambda$'s parameters as follows
\begin{equation}
{\mathbb B}\left( \theta \right) =\cosh \left( i\Lambda \right) +\frac{
\mbox{senh}\left( i\Lambda \right) }{\Lambda }\left( 
\begin{array}{cc}
\lambda _{3} & \lambda _{1} \\ 
\lambda _{2} & -\lambda _{3}
\end{array}
\right) . 
\end{equation}
Under the above considerations, the thermal operators are given by
\begin{eqnarray}
A\left( \theta \right) &=&\left[ \cosh \left( i\Lambda \right) +\frac{%
\lambda _{3}}{\Lambda }\mbox{sinh}\left( i\Lambda \right) \right] A+\frac{\lambda
_{1}}{\Lambda }\mbox{sinh}\left( i\Lambda \right) \widetilde{A}^{\dagger }
=u\left( \theta\right) A+v\left( \theta \right) \widetilde{A}^{\dagger }, \label{atran}
\\
\breve{A}\left( \theta \right) &=&\left[ \cosh \left( i\Lambda \right) +%
\frac{\lambda _{3}}{\Lambda }\mbox{sinh}\left( i\Lambda \right) \right] \widetilde{A%
}+\frac{\lambda _{1}}{\Lambda }\mbox{sinh}\left( i\Lambda \right) A^{\dagger }
=u\left( \theta \right) \widetilde{A}+v\left(\theta \right) A^{\dagger }, \label{attran}
\\
A^{\dagger }\left( \theta \right) &=&\left[ \cosh \left( i\Lambda \right) -%
\frac{\lambda _{3}}{\Lambda }\mbox{sinh}\left( i\Lambda \right) \right] A^{\dagger
}+\frac{\lambda _{2}}{\Lambda }\mbox{sinh}\left( i\Lambda \right) \widetilde{A}
=u^{*}\left( \theta \right) A^{\dagger }
+v^{*}\left( \theta\right) \widetilde{A}, 
\\
\breve{A}^{\dagger }\left( \theta \right) &=&\left[ \cosh \left(
i\Lambda \right) -\frac{\lambda _{3}}{\Lambda }\mbox{sinh}\left( i\Lambda \right)
\right] \widetilde{A}^{\dagger }+\frac{\lambda _{2}}{\Lambda }\mbox{sinh}\left(
i\Lambda \right) A
=u^{*}\left( \theta \right) \widetilde{A}^{\dagger }
+v^{*}\left( \theta \right) A.
\end{eqnarray}
In the above expressions, for thermal operators, we have an explicit evidence
that in fact $\breve{A}(\theta) \neq \widetilde{A}(\theta)$. To determine the
new rules, in the transformed space, we must execute the thermal Bogoliubov
transformation over the Tomita-Takesaki modular operator $J$:
\begin{equation}
J(\theta)=e^{-iG}Je^{iG}=e^{-iG}e^{-iG^{*}}J=Je^{iG^{*}}e^{iG},
\label{jtran}
\end{equation}
where we have used the property $JJ=1$ and the anti-linearity of $J$ {\cite{lvw}}.
Noting that $\left[G^{*}\right]^{\dagger}=G^{*}$, the expression (\ref{jtran})
clearly gives us that $J(\theta)=J^{\dagger}(\theta)$. Furthermore, it is easy to
show that $J(\theta)J(\theta)=1$. To see how this transformed operator works,
let's take the expression for $A(\theta)$ given at (\ref{atran}), and apply
the $J(\theta)$ over it:
\begin{equation}
J(\theta)A(\theta)J(\theta)=
\left(e^{-iG}e^{-iG^{*}}J\right)\left(e^{-iG}Ae^{iG}\right)\left(Je^{iG^{*}}e^{iG}\right),
\end{equation}
where we have made use of the expressions given at (\ref{jtran}). From the expression
above we have
\begin{equation}
J(\theta)A(\theta)J(\theta)=e^{-iG}JAJe^{iG}=e^{-iG}\widetilde{A}e^{iG}=\breve{A}(\theta),
\end{equation}
with $\breve{A}(\theta)$ given in (\ref{attran}).

The next step is to analise the result of the transformed Tomita-Takesaki modular
o\-pe\-ra\-tor actuation over the thermal vacuum.

The thermal vacuum and its thermal dual are obtained by using the $SU(1,1)$
``desentanglement'' theorem {\cite{eber,cha} and have the following condensed
state form 
\begin{equation}
\left| 0\left( \theta \right) \right\rangle =e^{iG\left( \theta \right) }
\left. \left| 0\right\rangle\!\right\rangle
=\frac{1}{u}e^{-\frac{v}{u}\widetilde{A}^{\dagger }A^{\dagger }}
\left. \left| 0\right\rangle\!\right\rangle,
\qquad
\left\langle 0\left( \theta \right) \right| =\left\langle\! \left\langle
0\right. \right| e^{iG\left( \theta \right) }=\left\langle
\!\left\langle 0\right. \right| \frac{1}{u^{*}}e^{-\frac{v^{*}}{u^{*}}\widetilde{A}A}, 
\label{tv}
\end{equation}
where, $\left. \left| 0\right\rangle\!\right\rangle = 
\left| 0\right\rangle \otimes \left|\tilde{0}\right\rangle$.
In this way we have the thermal vacuum definition 
\begin{equation}
A\left( \theta \right) \left| 0\left( \theta \right) \right\rangle =
\breve{A}\left( \theta \right) \left| 0\left( \theta \right)
\right\rangle =0,\qquad
\left\langle 0\left( \theta \right) \right| A^{\dagger }\left( \theta
\right) =\left\langle 0\left( \theta \right) \right| \breve{A}
^{\dagger }\left( \theta \right) =0.
\end{equation}
Furthermore, these states satisfy the thermal state conditions
\begin{eqnarray}
\left[ A+\frac{v}{u}\widetilde{A}^{\dagger }\right] \left| 0\left( \theta
\right) \right\rangle &=&0,
\quad \left[ \widetilde{A}+\frac{v}{u}A^{\dagger }\right] \left| 0\left( \theta
\right) \right\rangle =0, \\
\left\langle 0\left( \theta \right) \right| \left[ A^{\dagger }+
\frac{v^{*}}{u^{*}}\widetilde{A}\right] &=&0, \quad
\left\langle 0\left( \theta \right) \right| \left[ \widetilde{A}^{\dagger }+
\frac{v^{*}}{u^{*}}A\right] =0.
\end{eqnarray}
With the actuation $J(\theta)$, given in (\ref{jtran}), over the thermal vacuum
presented in (\ref{tv}), we have
\begin{equation}
J(\theta)\left| 0\left( \theta \right)
\right\rangle =\left(e^{-iG}e^{-iG^{*}}J\right)e^{-iG}
\left. \left| 0\right\rangle\!\right\rangle=e^{-iG}\left. \left| 0\right\rangle\!\right\rangle=
\left| 0\left( \theta \right)
\right\rangle.
\end{equation}
This means that the thermal vacuum is invariant under $J(\theta)$.
What we have now at hands is a completely well defined general unitary TFD
formulation, which can be used for inserting temperature into physical systems.
As the thermal Bogoliubov transformation is a canonical one, we have that all
the properties of the Tomita-Takesaki modular operator at zero temperature are
still valid at finite temperature.

Once established the general unitary $SU(1,1)$ TFD formulation, we can approach
the results presented at Ref. {\cite{agv2}} for bosonic strings and
bosonic $D_{p}$-branes at finite temperature, in order to expose the real
interpretation of those results.

We should say that the formulation presented in this letter was already been
carried out \cite{agv2}. However, in that work, the rules played by the Tomita-Takesaki
modular operator as well as its transformation for a temperature dependent system was
not completely accomplished. As a conclusion, the only necessary modification
is to change $\widetilde{A}(\beta) \rightarrow \breve{A}(\beta)$, where $\beta$ stands
for the inverse of the temperature and explicites the operators (and states) dependence
with respect to it. We would like to stress the fact that all the expressions in that
reference under question are correctly calculated, because they were obtained by using
the brute force, without making use of the tilde conjugation rules to map elements
from one space in those of the other space. Now, after this letter, all we must to do
is to exchange the symbol tilde by breve according to (\ref{tbu}) and (\ref{tbui}).
Clarified this fact, it is easy to check that under the action of the transformed
Tomita-Takesaki modular oparator, $J(\theta)$, the closed bosonic string thermal vacuum
and the thermal boundary state, interpreted as a bosonic $D_{p}$-brane at finite
temperature \cite{agv2}, remain invariant.

Note that once the transformation generator satisfies the properties as those
presented in this letter in expressions (\ref{gprop}), it is not necessary to
explicit its form in terms of oscillators-like operator of the doubled space to check
the main result of the work presented here: a consistent alternative general
TFD formulation. Formulation that, in principle, seems to take the advantages of the
$\alpha$-degree of freedom {\cite{ume, um3}} on one hand and uses a unitary
transformation, in the finite volume limite, on the other hand, stands as a suitable
formulation in order to deal with thermal states obtained from the action of thermal
operators over the thermal vacuum. That is the case of the system studied in
Ref. {\cite{agv2}}, for example.

In this letter, we discuss a generalization for the thermal Bogoliubov transformation,
the general $SU(1,1)$ unitary transformation, and show its consistency in the ambit of
Thermo Field Dynamics. By redefinition of the tilde conjugation rules, we show how to
use the general unitary $SU(1,1)$ TFD formulation without leading to ambiguous results,
providing a useful tool for the study of thermal effects as well as quantum dissipation
and information theory.  
  
\section*{Acknowledgements}
The authors would like to thanks Ion V. Vancea and Ademir E. Santana for useful
discussions. M. C. B. A. was supported by the CNPq grant 302019/2003-0, A. L. G.
was supported by a FAPESP post-doc fellowship.

\end{document}